\documentstyle[seceq,preprint]{ptptex}
\newcommand{\bra}[1]{\langle {#1} |}     %%
\newcommand{\ket}[1]{| {#1} \rangle}     %%
\newcommand{\dket}[1]{|\!| {#1} \rangle}     %%
\newcommand{\wtilde}[1]{\widetilde{#1}} %%
\newcommand{\wwtilde}[1]{\widetilde{\widetilde{#1}}} %%

\title{%        %You can use \\ for explicit line-break
Deformed Boson Scheme Stressing Even-Odd \\
Boson Number Difference. III
}
\subtitle{Parameter-Dependent Deformation
}    %use this when you want a subtitle

\author{%       %Use \sc for the family name
Atsushi {\sc Kuriyama},$^{1}$ 
Constan\c{c}a {\sc Provid\^encia},$^{2}$ \\
Jo\~ao da {\sc Provid\^encia},$^{2}$ Yasuhiko {\sc Tsue}$^{3}$ 
and Masatoshi {\sc Yamamura}$^{1}$
}

\inst{%         %Affiliation, neglected when [addenda] or [errata]
$^1$Faculty of Engineering, Kansai University, Suita 564-8680, Japan\\
$^{2}$Departamento de Fisica, Universidade de Coimbra, P-3000 Coimbra, 
Portugal\\
$^{3}$Physics Division, Faculty of Science, Kochi University, Kochi 780-8520, 
Japan
}

\recdate{%      %Editorial Office will fill in this.
\today
}

\abst{%       %this abstract is neglected when [addenda] or [errata]
A basic idea is proposed for extending the formalism of 
Part (II) of the series of our paper to the case of the 
parameter-dependent deformation. 
It is stressed that, through this extension, the variety 
of the application increases. Further, an answer for the 
problem mentioned in Part (II) is presented. 
}

\begin{document}

\maketitle

\section{Introduction}

In Part (II),\cite{one} the present authors proposed a possible 
form of the boson coherent state which consists of the 
orthogonal states with the even- and the odd-boson states. 
This is a superposition of the boson-pair states $\ket{ch}$ 
and $\ket{sh}$, which have been investigated in Part 
(I).\cite{two} 
The coefficients of the superposition are given by new parameters and 
the special case corresponds to the conventional 
boson coherent state. 
The basic viewpoint exists in the deformed boson scheme 
investigated by the present authors\cite{three} 
and 
the idea comes from that proposed also by the present 
authors.\cite{four} 
In Part (II), ${\wtilde f}(k)$ characterizing the deformation 
was of the parameter-independence. 
For the sake of this fact, the applicability of the deformed boson 
scheme presented in (II) may be not so wide as that expected. 

The main aim of this paper, Part (III), may be clear : 
The basic idea of Part (II) is extended to the case in which 
${\wtilde f}(k)$ is of the parameter-dependence. 
We can show that even if ${\wtilde f}(k)$ depends on the 
parameter, the essential part of the formalism given in (II) is 
unchanged. 
Only the relation (II.2$\cdot$9a), then, (II.2$\cdot$17), is 
slightly changed. 
Further, we will demonstrate the variety of the applications 
by adopting various types of ${\wtilde f}(k)$ which depend on 
the parameter. 

In \S 2, first, the incentive to the reformulation of Part (II) 
is mentioned, and then, the reformulation is performed. 
Section 3 is devoted to treating two illustrative cases. 
In \S\S 4 and 5, various concrete examples for the illustrative 
cases are discussed. 
Through this discussion, the variety of the applications 
is demonstrated. Finally, in \S 6, mainly, the problem mentioned 
in Part (II) is solved.

\section{Incentive to the reformulation of Part (II)}

Following Part (II), we continue to investigate the following state 
in the boson space constructed by the boson operator 
$({\hat c}, {\hat c}^*)$ :
\begin{eqnarray}
& &\ket{cr}=u\ket{ch}+(\gamma^*/|\gamma|)v\ket{sh}\ , 
\label{2-1}\\
& &\ket{ch}=\left(\sqrt{\Gamma_{ch}}\right)^{-1}\dket{ch} \ , \qquad
\ket{sh}=\left(\sqrt{\Gamma_{sh}}\right)^{-1}\dket{sh} \ , 
\label{2-2}
\end{eqnarray}
\vspace{-0.8cm}
\begin{subequations}\label{2-3}
\begin{eqnarray}
& &\dket{ch}=\ket{0}+\sum_{n=1}^{\infty} \frac{\gamma^{2n}}{\sqrt{(2n)!}}
{\wtilde f}(0){\wtilde f}(1)\cdots {\wtilde f}(2n-1)\ket{2n} \ , 
\label{2-3a}\\
& &\dket{sh}=\sum_{n=0}^{\infty} \frac{\gamma^{2n+1}}{\sqrt{(2n+1)!}}
{\wtilde f}(0){\wtilde f}(1)\cdots {\wtilde f}(2n)\ket{2n+1} \ , 
\label{2-3b}
\end{eqnarray}
\end{subequations}
\begin{equation}\label{2-4}
\ket{k}=\left(\sqrt{k!}\right)^{-1}({\hat c}^*)^k \ket{0} \ . \qquad
({\hat c}\ket{0}=0)
\end{equation}
Here, ${\wtilde f}(k)$ denotes function of $k$ which does not depend on any 
parameter : 
\begin{equation}\label{2-5}
{\wtilde f}(0)=1 \ , \qquad {\wtilde f}(k) > 0 \ . \qquad
(k=1,2,3,\cdots) 
\end{equation}
The quantities $\gamma$ and $v$ denote complex parameters and $u$ is real one 
obeying 
\begin{equation}\label{2-6}
u^2+|v|^2=1 \ . 
\end{equation}
The normalization constants $\Gamma_{ch}$ and $\Gamma_{sh}$ are given by 
\begin{subequations}\label{2-7}
\begin{eqnarray}
& &\Gamma_{ch}=1+\sum_{n=1}^{\infty}\frac{(|\gamma|^2)^{2n}}{(2n)!}
\left({\wtilde f}(0){\wtilde f}(1)\cdots {\wtilde f}(2n-1)\right)^2 \ , 
\label{2-7a}\\
& &\Gamma_{sh}=\sum_{n=0}^{\infty}\frac{(|\gamma|^2)^{2n+1}}{(2n+1)!}
\left({\wtilde f}(0){\wtilde f}(1)\cdots {\wtilde f}(2n)\right)^2 \ . 
\label{2-7b}
\end{eqnarray}
\end{subequations}
Under the above $\Gamma_{ch}$ and $\Gamma_{sh}$, $\ket{cr}$ is normalized as 
\begin{equation}\label{2-8}
\bra{cr}cr\rangle=1\ . 
\end{equation}

First, we must mention the incentive to the reformulation of Part (II). 
Let a certain state $\ket{cr}$ give us, for example, 
the following normalization constant : 
\begin{equation}\label{2-9}
\Gamma_{ch}=\sqrt{1+|\gamma|^4} \ , \qquad
\Gamma_{sh}=|\gamma|^2\left(\sqrt{1+|\gamma|^4}\right)^3 \ . 
\end{equation}
The above $\Gamma_{ch}$ and $\Gamma_{sh}$ can be formally expanded as 
\begin{subequations}\label{2-10}
\begin{eqnarray}
\Gamma_{ch}&=&1+\sum_{n=1}^{\infty}\frac{(|\gamma|^2)^{2n}}{(2n)!}
(-)^{n+1}(2n-3)!!(2n-1)!! \ , 
\label{2-10a}\\
\Gamma_{sh}&=&|\gamma|^2+(3/2)|\gamma|^6+(3/8)|\gamma|^{10} \nonumber\\
& &+3\sum_{n=3}^{\infty}\frac{(|\gamma|^2)^{2n+1}}{(2n+1)!}
(-)^n(2n-5)!!(2n+1)!! \ . 
\label{2-10b}
\end{eqnarray}
\end{subequations}
The form (\ref{2-10}) shows two troubles to be solved. 
One is related to the convergence of the series expansion. 
The expansion (\ref{2-10}) is divergent in the domain $|\gamma|^2\geq 1$. 
The other is more serious than the first. The cases $n=$even in the 
expansion (\ref{2-10a}) and $n=$odd in the expansion (\ref{2-10b}) give us 
negative expansion coefficients and this means that the states $\ket{2n}$ 
and $\ket{2n+1}$ cannot present us positive-definite norms. 
Therefore, inevitably, we have to conclude that there does not exist the 
state with the form (\ref{2-3}) which leads to the normalization constant 
(\ref{2-9}). 

In order to solve the above-mentioned trouble, we rewrite $\Gamma_{ch}$ and 
$\Gamma_{sh}$ as follows : 
\begin{equation}\label{2-11}
\Gamma_{ch}=\left(\sqrt{1-\frac{|\gamma|^4}{1+|\gamma|^4}}\right)^{-1} \ , 
\quad
\Gamma_{sh}=|\gamma|^2
\left(\sqrt{1-\frac{|\gamma|^4}{1+|\gamma|^4}}\right)^{-3} \ . 
\end{equation}
Then, $\Gamma_{ch}$ and $\Gamma_{sh}$ can be expanded in terms of 
$|\gamma|^4/(1+|\gamma|^4)$ which satisfies $|\gamma|^4/(1+|\gamma|^4)<1$ 
for any magnitude of $|\gamma|^4$ : 
\begin{subequations}\label{2-12}
\begin{eqnarray}
& &\Gamma_{ch}=1+\sum_{n=1}^{\infty}\frac{(|\gamma|^2)^{2n}}{(2n)!}
[(2n-1)!!]^2\frac{1}{(1+|\gamma|^4)^n} \ , 
\label{2-12a}\\
& &\Gamma_{sh}=\sum_{n=0}^{\infty}\frac{(|\gamma|^2)^{2n+1}}{(2n+1)!}
[(2n+1)!!]^2\frac{1}{(1+|\gamma|^4)^n} \ . 
\label{2-12b}
\end{eqnarray}
\end{subequations}
The states $\dket{ch}$ and $\dket{sh}$ which give the above $\Gamma_{ch}$ and 
$\Gamma_{sh}$ are obtained in the form 
\begin{subequations}\label{2-13}
\begin{eqnarray}
& &\dket{ch}=\ket{0}+\sum_{n=1}^{\infty}\frac{\gamma^{2n}}{\sqrt{(2n)!}}
(2n-1)!!\frac{1}{(\sqrt{1+|\gamma|^4})^n}\ket{2n} \ , 
\label{2-13a}\\
& &\dket{sh}=\sum_{n=0}^{\infty}\frac{\gamma^{2n+1}}{\sqrt{(2n+1)!}}
(2n+1)!!\frac{1}{(\sqrt{1+|\gamma|^4})^n}\ket{2n+1} \ . 
\label{2-13b}
\end{eqnarray}
\end{subequations}
We can see that, in the relations (\ref{2-12}) and (\ref{2-13}), 
the trouble mentioned in the part after the relation (\ref{2-10} 
is completely removed. 
However, associating with this fact, we must understand that the 
function ${\wtilde f}(k)$ characterizing the deformation depends on 
the real parameter $|\gamma|^4$. 
In this case, ${\wtilde f}(k)$ is given as 
\begin{equation}\label{2-14}
{\wtilde f}(2n-1)=\frac{1}{\sqrt{1+|\gamma|^4}} \ , \qquad
{\wtilde f}(2n)=2n+1 \ . 
\end{equation}
Thus, we have to conclude that, in order to treat the case 
such as the form (\ref{2-9}) in the framework of Part (II), the reformulation 
is necessary so as to be able to describe the case in which ${\wtilde f}(k)$ 
depends on real parameter.

Following the above-mentioned incentive, we reformulate Part (II). 
This task is quite easily performed. 
All the relations except for the relation (II.2$\cdot$9a), and 
then (II.2$\cdot$17) are still available in the present case. 
In the case where ${\wtilde f}(k)$ depends on real parameter, 
$\Gamma_{ch}$ and $\Gamma_{sh}$ are expressed as 
\begin{subequations}\label{2-15}
\begin{eqnarray}
& &\Gamma_{ch}=1+\sum_{n=1}^{\infty}\frac{(|\gamma|^2)^{2n}}{(2n)!}
\left({\wtilde f}(0,\delta){\wtilde f}(1,\delta)\cdots{\wtilde f}(2n-1,\delta)
\right)^2 \ , 
\label{2-15a}\\
& &\Gamma_{sh}=\sum_{n=0}^{\infty}\frac{(|\gamma|^2)^{2n+1}}{(2n+1)!}
\left({\wtilde f}(0,\delta){\wtilde f}(1,\delta)\cdots{\wtilde f}(2n,\delta)
\right)^2 \ . 
\label{2-15b}
\end{eqnarray}
\end{subequations}
Here, $\delta$ denotes real parameter including $|\gamma|^4$ and 
${\wtilde f}(k,\delta)$, of course, should satisfy 
\begin{equation}\label{2-16}
{\wtilde f}(0,\delta)=1 \ , \qquad
{\wtilde f}(k,\delta)>0 \ . \quad (k=1,2,3,\cdots)
\end{equation}
For the derivative of $\Gamma_{ch}$ and $\Gamma_{sh}$ with respect to 
$|\gamma|^2$ are defined as 
\begin{subequations}\label{2-17}
\begin{eqnarray}
& &\Gamma_{ch}'=\sum_{n=1}^{\infty}\frac{1}{(2n)!}\frac{d}{d|\gamma|^2}
(|\gamma|^2)^{2n}\cdot
\left({\wtilde f}(0,\delta){\wtilde f}(1,\delta)\cdots{\wtilde f}(2n-1,\delta)
\right)^2 \ , 
\label{2-17a}\\
& &\Gamma_{sh}'=\sum_{n=0}^{\infty}\frac{1}{(2n+1)!}
\frac{d}{d|\gamma|^2}{(|\gamma|^2)^{2n+1}}\cdot
\left({\wtilde f}(0,\delta){\wtilde f}(1,\delta)\cdots{\wtilde f}(2n,\delta)
\right)^2 \ . 
\label{2-17b}
\end{eqnarray}
\end{subequations}
It should be noted that, even if $\delta=|\gamma|^4$ or a function of 
$|\gamma|^4$, the differential of ${\wtilde f}(k,\delta)$ for $|\gamma|^2$ 
is not necessary to calculate. 
We can expect that the variety in the application of the basic idea of 
Part (II) increases if we accept the parameter-dependent ${\wtilde f}(k)$.

\section{Two illustrative cases}

In order to demonstrate the usefulness of the idea 
given in \S 2, first, we investigate the following case : 
\begin{subequations}\label{3-1}
\begin{eqnarray}
& &\dket{ch}=\ket{0}+\sum_{n=1}^{\infty}\frac{\gamma^{2n}}{\sqrt{(2n)!}}
(2n-1)!!(g_{ch})^n\ket{2n} \ , 
\label{3-1a}\\
& &\dket{sh}=\sum_{n=0}^{\infty}\frac{\gamma^{2n+1}}{\sqrt{(2n+1)!}}
(2n+1)!!(g_{sh})^n\ket{2n+1} \ . 
\label{3-1b}
\end{eqnarray}
\end{subequations}
Here, $g_{ch}$ and $g_{sh}$ denote positive functions for 
$|\gamma|^4$. The normalization constants $\Gamma_{ch}$ and $\Gamma_{sh}$ 
are given in the form 
\begin{subequations}\label{3-2}
\begin{eqnarray}
& &\Gamma_{ch}=1+\sum_{n=1}^{\infty}\frac{(|\gamma|^2)^{2n}}{{(2n)!}}
[(2n-1)!!]^2(g_{ch})^{2n} \ , 
\label{3-2a}\\
& &\Gamma_{sh}=\sum_{n=0}^{\infty}\frac{(|\gamma|^2)^{2n+1}}{{(2n+1)!}}
[(2n+1)!!]^2(g_{sh})^{2n} \ . 
\label{3-2b}
\end{eqnarray}
\end{subequations}
The expansion (\ref{3-2}) is convergent, if $g_{ch}$ and $g_{sh}$ obey 
\begin{equation}\label{3-3}
|\gamma|^4g_{ch}^2 <1 \ , \qquad 
|\gamma|^4 g_{sh}^2 <1 \ . 
\end{equation}
Under this condition, the expansion (\ref{3-2}) can be expressed compactly 
as 
\begin{equation}\label{3-4}
\Gamma_{ch}=\left(\sqrt{1-|\gamma|^4g_{ch}^2}\right)^{-1} \ , \qquad
\Gamma_{sh}=|\gamma|^2\left(\sqrt{1-|\gamma|^4g_{sh}^2}\right)^{-3} \ . 
\end{equation}
The case $g_{ch}^2=g_{sh}^2=(1+|\gamma|^4)^{-1}$ corresponds to the relation 
(\ref{2-11}). The relation (\ref{2-17}) gives 
\begin{subequations}\label{3-5}
\begin{eqnarray}
& &\Gamma_{ch}'
=|\gamma|^2g_{ch}^2\left(\sqrt{1-|\gamma|^4g_{ch}^2}\right)^{-3} 
\ , \label{3-5a}\\
& &\Gamma_{sh}'=\left(\sqrt{1-|\gamma|^4g_{sh}^2}\right)^{-3} 
+3|\gamma|^4g_{sh}^2\left(\sqrt{1-|\gamma|^4g_{sh}^2}\right)^{-5} 
\ . \label{3-5b}
\end{eqnarray}
\end{subequations}
Then, we have 
\begin{subequations}\label{3-6}
\begin{eqnarray}
& &\left(\frac{\Gamma'}{\Gamma}\right)_{ch}
=\frac{|\gamma|^2g_{ch}^2}{1-|\gamma|^4g_{ch}^2} \ , 
\label{3-6a}\\
& &\left(\frac{\Gamma'}{\Gamma}\right)_{sh}-\frac{1}{|\gamma|^2}
=\frac{3|\gamma|^2g_{sh}^2}{1-|\gamma|^4g_{sh}^2} \ . 
\label{3-6b}
\end{eqnarray}
\end{subequations}
The form (\ref{3-6}) leads to 
\begin{equation}\label{3-7}
2|c|^2=|\gamma|^4\left(\frac{u^2 g_{ch}^2}{1-|\gamma|^4 g_{ch}^2}
+\frac{3|v|^2 g_{sh}^2}{1-|\gamma|^4 g_{sh}^2}\right) \ .
\end{equation}
Here, we used the relation (II.2$\cdot$17). We can determine 
$|\gamma|^4$ as a function of $2|c|^2$, if $g_{ch}$ and $g_{sh}$ are given 
concretely. 
Then, following the method shown in Part (II), $\gamma$ and $v$ can be 
expressed in the form 
\begin{equation}\label{3-8}
\gamma=\sqrt{\sqrt{2}c}\cdot \sqrt[4]{F_{cr}} \ , \qquad v=\eta \ . 
\end{equation}
The functions ${\wtilde f}(2n-1,\delta)$ and ${\wtilde f}(2n,\delta)$ 
in the present case can be calculated as 
\begin{equation}\label{3-9}
{\wtilde f}(2n-1,\delta)=g_{ch}\cdot (g_{ch}/g_{sh})^{n-1} \ , \quad
{\wtilde f}(2n,\delta)=(2n+1)\cdot (g_{sh}/g_{ch})^n \ . 
\end{equation}
With the use of the form (\ref{3-9}), the expectation value of 
${wtilde f}({\hat N})\cdot{\wtilde f}({\hat N}+1)$ for $\ket{ch}$ and 
$\ket{sh}$ is obtained : 
\begin{subequations}\label{3-10}
\begin{eqnarray}
& &\left({\wtilde f}(N)\cdot{\wtilde f}(N+1)\right)_{ch}
=\frac{g_{ch}}{1-|\gamma|^4 g_{ch}^2} \ , 
\label{3-10a}\\
& &\left({\wtilde f}(N)\cdot{\wtilde f}(N+1)\right)_{sh}
=\frac{3g_{sh}}{1-|\gamma|^4 g_{sh}^2} \ . 
\label{3-10b}
\end{eqnarray}
\end{subequations}
These are used for calculating the expectation value of ${\hat c}^2/2$.

Next, we investigate the following case : 
\begin{subequations}\label{3-11}
\begin{eqnarray}
& &\dket{ch}=\ket{0}+\sum_{n=1}^{\infty}\frac{\gamma^{2n}}{\sqrt{(2n)!}}
\sqrt{(2n-1)!!}(g_{ch})^n\ket{2n} \ , 
\label{3-11a}\\
& &\dket{sh}=\sum_{n=0}^{\infty}\frac{\gamma^{2n+1}}{\sqrt{(2n+1)!}}
\sqrt{(2n+1)!!}(\sqrt{3}g_{sh})^n\ket{2n+1} \ . 
\label{3-11b}
\end{eqnarray}
\end{subequations}
The above form should be compared with the form (\ref{3-1}). 
The normalization constants are given as 
\begin{subequations}\label{3-12}
\begin{eqnarray}
& &\Gamma_{ch}=1+\sum_{n=1}^{\infty}\frac{(|\gamma|^2)^{2n}}{{(2n)!}}
(2n-1)!!(g_{ch})^{2n} =\exp\left(|\gamma|^4 g_{ch}^2/2\right) \ , 
\label{3-12a}\\
& &\Gamma_{sh}=\sum_{n=0}^{\infty}\frac{(|\gamma|^2)^{2n+1}}{{(2n+1)!}}
(2n+1)!!(\sqrt{3}g_{sh})^{2n} 
=|\gamma|^2\exp\left(3|\gamma|^4 g_{sh}^2/2\right) \ . 
\label{3-12b}
\end{eqnarray}
\end{subequations}
The quantities $\Gamma_{ch}'$ and $\Gamma_{sh}'$ are calculated as 
\begin{subequations}\label{3-13}
\begin{eqnarray}
& &\Gamma_{ch}'
=|\gamma|^2g_{ch}^2\exp\left(|\gamma|^4g_{ch}^2/2\right) \ , 
\label{3-13a}\\
& &\Gamma_{sh}'=\left(1+3|\gamma|^4g_{sh}^2\right) 
\exp\left(3|\gamma|^4g_{sh}^2/2\right) \ . 
\label{3-13b}
\end{eqnarray}
\end{subequations}
Then, we have 
\begin{subequations}\label{3-14}
\begin{eqnarray}
& &\left(\frac{\Gamma'}{\Gamma}\right)_{ch}
=|\gamma|^2g_{ch}^2 \ , 
\label{3-14a}\\
& &\left(\frac{\Gamma'}{\Gamma}\right)_{sh}-\frac{1}{|\gamma|^2}
=3|\gamma|^2g_{sh}^2 \ . 
\label{3-14b}
\end{eqnarray}
\end{subequations}
With the use of the relation (\ref{3-14}), $2|c|^2$ is expressed in the form 
\begin{equation}\label{3-15}
2|c|^2=|\gamma|^4 (u^2g_{ch}^2+3|v|^2 g_{sh}^2 )\ . 
\end{equation}
In this case, also, we have 
\begin{equation}\label{3-16}
\gamma=\sqrt{\sqrt{2}c}\cdot \sqrt[4]{F_{cr}} \ , \qquad v=\eta \ .
\end{equation}
The functions ${\wtilde f}(2n-1,\delta)$ and ${wtilde f}(2n,\delta)$ in the 
present case can be given as 
\begin{subequations}\label{3-17}
\begin{eqnarray}
& &{\wtilde f}(2n-1,\delta)=g_{ch}(g_{ch}/\sqrt{3}g_{sh})^{n-1} \ , 
\label{3-17a}\\
& &{\wtilde f}(2n,\delta)=\sqrt{2n+1}\cdot(\sqrt{3}g_{sh}/g_{ch})^{n} \ . 
\label{3-17b}
\end{eqnarray}
\end{subequations}
With the use of the relation (\ref{3-17}), we can calculate the 
quantities 
$({\wtilde f}({N})\cdot{\wtilde f}({N}+1))_{ch}$ and 
$({\wtilde f}({N})\cdot{\wtilde f}({N}+1))_{sh}$. 
However, the existence of $\sqrt{2n+1}$ in the relation (\ref{3-17b}) 
does not give the exact result in compact form : 
\begin{subequations}\label{3-18}
\begin{eqnarray}
& &\left({\wtilde f}(N)\cdot{\wtilde f}(N+1)\right)_{ch}
\approx g_{ch}\sqrt{1+|\gamma|^4 g_{ch}^2} \ , 
\label{3-18a}\\
& &\left({\wtilde f}(N)\cdot{\wtilde f}(N+1)\right)_{sh}
\approx 3g_{sh}\sqrt{1+|\gamma|^4 g_{sh}^2} \ . 
\label{3-18b}
\end{eqnarray}
\end{subequations}
The above approximate expressions are obtained under the idea presented 
in Part (I).

\section{The simplest examples}

With the use of various examples, we can investigate the relations presented 
in \S 3 in more detail. First example is the following case for the state 
(\ref{3-1a}) and (\ref{3-1b}) :
\begin{equation}\label{4-1}
g_{ch}=\sqrt{\frac{\lambda}{1+\lambda|\gamma|^4}}\ , \qquad
g_{sh}=\sqrt{\frac{\mu}{1+\mu|\gamma|^4}}\ . 
\end{equation}
Here, $\lambda$ and $\mu$ denote positive constants. 
Then, the relation (\ref{3-7}) is reduced to 
\begin{eqnarray}\label{4-2}
& &2|c|^2=|\gamma|^4(u^2 \lambda + 3|v|^2\mu) \ , \nonumber\\
\hbox{\rm i.e.,}\qquad & & \nonumber\\
& &|\gamma|^4=2|c|^2(u^2\lambda + 3|v|^2\mu)^{-1} \ . 
\end{eqnarray}
The above relation gives 
\begin{equation}\label{4-3}
\gamma=\sqrt{\sqrt{2}c}\cdot \left(\sqrt[4]{u^2\lambda+3|v|^2\mu}\right)^{-1} 
\ .
\end{equation}
The expectation value of ${\hat \tau}_-$ which is defined in (II.2$\cdot$19) 
is calculated in the form 
\begin{eqnarray}\label{4-4}
(\tau_-)_{cr}&=&\frac{u^2\lambda}{u^2\lambda+3|v|^2\mu} c 
\sqrt{\frac{u^2\lambda+3|v|^2\mu}{2\lambda}+|c|^2} \nonumber\\
& &+\frac{3|v|^2\mu}{u^2\lambda+3|v|^2\mu} c 
\sqrt{\frac{u^2\lambda+3|v|^2\mu}{2\mu}+|c|^2} \ , \nonumber\\
u^2&=&1-|\eta|^2 \ , \qquad |v|^2=|\eta|^2 \ . 
\end{eqnarray}
The quantity $(\tau_0)_{cr}$ is given in the relation (II.2$\cdot$21) : 
\begin{equation}\label{4-5}
(\tau_0)_{cr}=|c|^2+|\eta|^2/2+1/4 \ . 
\end{equation}
The cases $\lambda=\mu$ and $\lambda=3\mu$ lead to the following forms, 
respectively : 
\begin{eqnarray}
& &(\tau_-)_{cr}=c\sqrt{1/2+|\eta|^2+|c|^2} \ , \qquad (\lambda=\mu)
\label{4-6}\\
& &(\tau_-)_{cr}=u^2c\sqrt{1/2+|c|^2} +|v|^2 c\sqrt{3/2+|c|^2} 
\ , \qquad (\lambda=3\mu)
\label{4-7}
\end{eqnarray}
The forms (\ref{4-6}) and (\ref{4-7}) are identical with the relations 
(II.3$\cdot$10) and (II.3$\cdot$20), respectively. 
The case (II.3$\cdot$20) is approximately obtained. Further, if 
$\lambda \rightarrow \infty$ ($\mu$ is finite) and $\mu \rightarrow \infty$ 
($\lambda$ is finite), we have, respectively, 
\begin{eqnarray}
& &(\tau_-)_{cr}=c\sqrt{u^2/2+|c|^2} \ , \qquad (\lambda\rightarrow \infty)
\label{4-8}\\
& &(\tau_-)_{cr}=c\sqrt{|v|^2/2+|c|^2}\ , \qquad (\mu\rightarrow \infty)
\label{4-9}
\end{eqnarray}
From the above argument, we can understand that, by changing the magnitudes 
of $\lambda$ and $\mu$, infinite types for the deformations are derived. 
This is one of the merits for adopting parameter-dependent ${\wtilde f}(k)$. 
The case shown in the introductory part of \S 2 corresponds to the case 
$\lambda=\mu$.

Next, we treat the following case for the states (\ref{3-11a}) and 
(\ref{3-11b}) : 
\begin{equation}\label{4-10}
g_{ch}=\sqrt{\lambda} \ , \qquad g_{sh}=\sqrt{\mu} \ . 
\end{equation}
Here, of course, $\lambda$ and $\mu$ are positive constants. 
Therefore, the relation (\ref{3-15}) becomes of the same form as the form 
(\ref{4-2}), and then, the form (\ref{4-3}). 
This means that the relations (\ref{4-4}) and (\ref{4-5}) keep their forms 
in the present case. 
However, the relation (\ref{4-4}) holds approximately.

\section{More complicated examples}

Our next investigation concerns with more complicate example than 
those discussed in \S 4. 
For the state (\ref{3-1}), we treat $g_{ch}$ and $g_{sh}$ in the following 
three cases : 
\begin{subequations}\label{5-1}
\begin{eqnarray}
\hbox{\rm (i)}\qquad 
& &g_{ch}=\sqrt{\frac{\lambda}
{\beta+\sqrt{\alpha^2+|\gamma|^4}+\lambda|\gamma|^4}} \ , 
\label{5-1a}\\
& &g_{sh}=\sqrt{\frac{\mu}
{\alpha+\sqrt{\alpha^2+|\gamma|^4}+\mu|\gamma|^4}} \ , 
\label{5-1b}
\qquad\qquad\qquad\qquad\qquad\qquad\qquad\qquad
\end{eqnarray}
\end{subequations}
\begin{subequations}\label{5-2}
\begin{eqnarray}
\hbox{\rm (ii)}\qquad 
& &g_{ch}=\sqrt{\frac{\lambda}
{\alpha+\sqrt{\alpha^2+|\gamma|^4}+\lambda|\gamma|^4}} \ , 
\label{5-2a}\\
& &g_{sh}=\sqrt{\frac{\mu}
{\beta+\sqrt{\alpha^2+|\gamma|^4}+\mu|\gamma|^4}} \ , 
\label{5-2b}
\qquad\qquad\qquad\qquad\qquad\qquad\qquad\qquad
\end{eqnarray}
\end{subequations}
\begin{subequations}\label{5-3}
\begin{eqnarray}
\hbox{\rm (iii)}\qquad 
& &g_{ch}=\sqrt{\frac{\lambda}
{\alpha+\sqrt{\alpha^2+|\gamma|^4}+\lambda|\gamma|^4}} \ , 
\label{5-3a}\\
& &g_{sh}=\sqrt{\frac{\mu}
{\beta+\sqrt{\beta^2+|\gamma|^4}+\mu|\gamma|^4}} \ , 
\label{5-3b}
\qquad\qquad\qquad\qquad\qquad\qquad\qquad\qquad
\end{eqnarray}
\end{subequations}
Here, $\alpha$, $\beta$, $\lambda$ and $\mu$ denote positive 
or negative constants. By substituting the above three forms into 
the relation (\ref{3-7}), we have the relations 
\begin{eqnarray}
& &2|c|^2=|\gamma|^4\left(
\frac{u^2\lambda}{\beta+\sqrt{\alpha^2+|\gamma|^4}}+
\frac{3|v|^2\mu}{\alpha+\sqrt{\alpha^2+|\gamma|^4}}\right) \ , 
\label{5-4}\\
& &2|c|^2=|\gamma|^4\left(
\frac{u^2\lambda}{\alpha+\sqrt{\alpha^2+|\gamma|^4}}+
\frac{3|v|^2\mu}{\beta+\sqrt{\alpha^2+|\gamma|^4}}\right) \ , 
\label{5-5}\\
& &2|c|^2=|\gamma|^4\left(
\frac{u^2\lambda}{\alpha+\sqrt{\alpha^2+|\gamma|^4}}+
\frac{3|v|^2\mu}{\beta+\sqrt{\beta^2+|\gamma|^4}}\right) \ , 
\label{5-6}
\end{eqnarray}
Although we do not show the explicit forms, the relations (\ref{5-4}) $\sim$ 
(\ref{5-6}) can be solved for $|\gamma|^4$ as functions of $2|c|^2$ in the 
framework of quadratics. 
Further, the relation (\ref{3-10}) is, 
in the present case, reduced to 
\begin{subequations}\label{5-7}
\begin{eqnarray}
& &\left({\wtilde f}(N)\cdot{\wtilde f}(N+1)\right)_{ch}
=\sqrt{\lambda}\sqrt{\frac{1}{\beta+\sqrt{\alpha^2+|\gamma|^4}}+
\frac{\lambda|\gamma|^4}{(\beta+\sqrt{\alpha^2+|\gamma|^4})^2}} \ , 
\label{5-7a}\qquad\\
& &\left({\wtilde f}(N)\cdot{\wtilde f}(N+1)\right)_{sh}
=3\sqrt{\mu}\sqrt{\frac{1}{\alpha+\sqrt{\alpha^2+|\gamma|^4}}+
\frac{\mu|\gamma|^4}{(\alpha+\sqrt{\alpha^2+|\gamma|^4})^2}} \ , 
\label{5-7b}
\end{eqnarray}
\end{subequations}
\begin{subequations}\label{5-8}
\begin{eqnarray}
& &\left({\wtilde f}(N)\cdot{\wtilde f}(N+1)\right)_{ch}
=\sqrt{\lambda}\sqrt{\frac{1}{\alpha+\sqrt{\alpha^2+|\gamma|^4}}+
\frac{\lambda|\gamma|^4}{(\alpha+\sqrt{\alpha^2+|\gamma|^4})^2}} \ , 
\label{5-8a}\qquad\\
& &\left({\wtilde f}(N)\cdot{\wtilde f}(N+1)\right)_{sh}
=3\sqrt{\mu}\sqrt{\frac{1}{\beta+\sqrt{\alpha^2+|\gamma|^4}}+
\frac{\mu|\gamma|^4}{(\beta+\sqrt{\alpha^2+|\gamma|^4})^2}} \ , 
\label{5-8b}
\end{eqnarray}
\end{subequations}
\begin{subequations}\label{5-9}
\begin{eqnarray}
& &\left({\wtilde f}(N)\cdot{\wtilde f}(N+1)\right)_{ch}
=\sqrt{\lambda}\sqrt{\frac{1}{\alpha+\sqrt{\alpha^2+|\gamma|^4}}+
\frac{\lambda|\gamma|^4}{(\alpha+\sqrt{\alpha^2+|\gamma|^4})^2}} \ , 
\label{5-9a}\qquad\\
& &\left({\wtilde f}(N)\cdot{\wtilde f}(N+1)\right)_{sh}
=3\sqrt{\mu}\sqrt{\frac{1}{\beta+\sqrt{\beta^2+|\gamma|^4}}+
\frac{\mu|\gamma|^4}{(\beta+\sqrt{\beta^2+|\gamma|^4})^2}} \ . 
\label{5-9b}
\end{eqnarray}
\end{subequations}
We can see that $(\alpha=1, \beta=0, \lambda=1, \mu=1/3)$ and 
$(\alpha=3/2, \beta=-1/2, \lambda=1, \mu=1/3)$ in the relation (\ref{5-4}) 
correspond to the relations (II.6$\cdot$7) and (II.6$\cdot$9), respectively.

First of all, we show an exact solution in a simple form ; the case 
$\alpha=\beta$. 
In this case, three forms (\ref{5-1}) $\sim$ (\ref{5-3}) are identical to 
each other. The relations (\ref{5-4}) $\sim$ (\ref{5-6}) are reduced to the 
following : 
\begin{equation}\label{5-10}
2|c|^2=|\gamma|^4\left(
\frac{u^2\lambda}{\alpha+\sqrt{\alpha^2+|\gamma|^4}}+
\frac{3|v|^2\mu}{\alpha+\sqrt{\alpha^2+|\gamma|^4}}\right) \ . 
\end{equation}
Further, the relations (\ref{5-7}) $\sim$ (\ref{5-9}) are reduced to 
\begin{subequations}\label{5-11}
\begin{eqnarray}
& &\left({\wtilde f}(N)\cdot{\wtilde f}(N+1)\right)_{ch}
=\sqrt{\lambda}\sqrt{\frac{1}{\alpha+\sqrt{\alpha^2+|\gamma|^4}}+
\frac{\lambda|\gamma|^4}{(\alpha+\sqrt{\alpha^2+|\gamma|^4})^2}} \ , 
\label{5-11a}\qquad\\
& &\left({\wtilde f}(N)\cdot{\wtilde f}(N+1)\right)_{sh}
=3\sqrt{\mu}\sqrt{\frac{1}{\alpha+\sqrt{\alpha^2+|\gamma|^4}}+
\frac{\mu|\gamma|^4}{(\alpha+\sqrt{\alpha^2+|\gamma|^4})^2}} \ . 
\label{5-11b}
\end{eqnarray}
\end{subequations}
The relation (\ref{5-10}) gives the following : 
\begin{eqnarray}
& &|\gamma|^4=2|c|^2\frac{2\alpha(u^2 \lambda + 3|v|^2\mu)+2|c|^2}
{(u^2\lambda+3|v|^2\mu)^2} \ , 
\label{5-12}\\
\hbox{\rm i.e.,}\qquad & & \nonumber\\
& &\gamma=\sqrt{\sqrt{2}c}\cdot \sqrt[4]
{\frac{2\alpha(u^2 \lambda + 3|v|^2\mu)+2|c|^2}
{(u^2\lambda+3|v|^2\mu)^2}} \ . 
\label{5-13}
\end{eqnarray}
The relation (\ref{5-11}) is expressed as 
\begin{subequations}\label{5-14}
\begin{eqnarray}
& &\left({\wtilde f}(N)\cdot{\wtilde f}(N+1)\right)_{ch}
=\lambda\sqrt{\frac{\frac{u^2\lambda+3|v|^2\mu}{2\lambda}+|c|^2}
{\alpha(u^2\lambda+3|v|^2\mu)+|c|^2}}
\ , 
\label{5-14a}\qquad\\
& &\left({\wtilde f}(N)\cdot{\wtilde f}(N+1)\right)_{sh}
=\mu\sqrt{\frac{\frac{u^2\lambda+3|v|^2\mu}{2\mu}+|c|^2}
{\alpha(u^2\lambda+3|v|^2\mu)+|c|^2}}
\ . 
\label{5-14b}
\end{eqnarray}
\end{subequations}
With the use of the relations (\ref{5-13}) and (\ref{5-14}), 
$(\tau_-)_{cr}$ is obtained in the form 
\begin{eqnarray}\label{5-15}
(\tau_-)_{cr}&=&\frac{u^2\lambda}{u^2\lambda+3|v|^2\mu} c 
\sqrt{\frac{u^2\lambda+3|v|^2\mu}{2\lambda}+|c|^2} \nonumber\\
& &+\frac{3|v|^2\mu}{u^2\lambda+3|v|^2\mu} c 
\sqrt{\frac{u^2\lambda+3|v|^2\mu}{2\mu}+|c|^2} \ , \nonumber\\
u^2&=&1-|\eta|^2 \ , \qquad |v|^2=|\eta|^2 \ . 
\end{eqnarray}
We can see that the form (\ref{5-15}) is identical to the relation (\ref{4-4}).

Next, we investigate the case $\alpha \neq \beta$. In principle, 
we can show the exact result in concrete form for this case. 
However, it is too complicated to demonstrate the physical significance. 
Then, we discuss the approximate forms for the tow regions 
$|\gamma|^2 \rightarrow \infty$ and $|\gamma|^2 \sin 0$. 
In the region $|\gamma|^2 \rightarrow \infty$, by picking the leading terms 
for $|\gamma|^4$ in the relation (\ref{5-4}), $F_{cr}$ introduced in the 
relation (II.2$\cdot$12) in the case (i) is given as 
\begin{equation}\label{5-16}
F_{cr}=\frac{2(|c|^2+(u^2\lambda\beta+3|v|^2\mu\alpha))}
{(u^2\lambda+3|v|^2\mu)^2} \ . 
\end{equation}
By exchanging $\alpha$ and $\beta$ in the form (\ref{5-16}), we obtain 
$F_{cr}$ for the case (ii) and (iii). 
Picking also the leading terms for $|\gamma|^4$ in the 
relation (\ref{5-7}), we have the following form for the case (i) : 
\begin{eqnarray}\label{5-17}
& &u^2\left({\wtilde f}(N)\cdot{\wtilde f}(N+1)\right)_{ch}
+|v|^2\left({\wtilde f}(N)\cdot{\wtilde f}(N+1)\right)_{sh} \nonumber\\
&=&u^2\lambda\sqrt{1+\left(\frac{1}{\lambda}-2\beta\right)\frac{
u^2\lambda+3|v|^2\mu}{2|c|^2}} \nonumber\\
& &+3|v|^2\mu\sqrt{1+\left(\frac{1}{\mu}-2\alpha\right)\frac{
u^2\lambda+3|v|^2\mu}{2|c|^2}} ={\wwtilde f} \ .
\end{eqnarray}
Since $2|c|^2$ is sufficiently large, ${\wtilde {\wtilde f}}$ can be 
rewritten as follows : 
\begin{equation}\label{5-18}
{\wwtilde{f}}
=(u^2\lambda+3|v|^2\mu)\sqrt{1-\frac{u^2(2\beta\lambda-1)
+3|v|^2(2\alpha\mu-1)}{2|c|^2}} \ .
\end{equation}
By exchanging $\alpha$ and $\beta$ in the forms (\ref{5-17}) and (\ref{5-18}), 
we obtain ${\wwtilde f}$ in the cases (ii) and (iii). 
With the use of the relations (\ref{5-16}) and (\ref{5-18}), 
$(\tau_-)_{cr}$ is obtained in the form 
\begin{eqnarray}\label{5-19}
(\tau_-)_{cr}&=&\frac{1}{2}\gamma^2\cdot{\wwtilde f} = 
c\sqrt{F_{cr}/2}\cdot{\wwtilde f} \nonumber\\
&=&c\sqrt{|c|^2+|v|^2+1/2} \ .
\end{eqnarray}
The form (\ref{5-19}) is also valid for the cases (ii) and (iii). 
It may be interesting to see that the form (\ref{5-19}) does not depend 
on the parameters $\alpha$, $\beta$, $\lambda$ and $\mu$.

On the other hand, the expressions in the region $|\gamma|^2 \sim 0$ are 
rather complicated. The quantity $F_{cr}$ for the cases (i), (ii) and (iii) 
are given in the following forms : 
\begin{eqnarray}
\hbox{\rm (i)}\quad
F_{cr}&=&
\left(\frac{u^2\lambda}{\alpha+\beta}+\frac{3|v|^2\mu}{2\alpha}\right)^{-1} 
\nonumber\\
& &\times 
\left[
1+\left(\frac{u^2\lambda}{2\alpha(\alpha+\beta)^2}+\frac{3|v|^2\mu}
{(2\alpha)^3}
\right)\left(\frac{u^2\lambda}{\alpha+\beta}+\frac{3|v|^2\mu}{2\alpha}
\right)^{-2}\cdot 2|c|^2\right] \ , \qquad\ \ 
\label{5-20}\\
\hbox{\rm (ii)}\quad
F_{cr}&=&
\left(\frac{u^2\lambda}{2\alpha}+\frac{3|v|^2\mu}{\alpha+\beta}\right)^{-1} 
\nonumber\\
& &\times 
\left[
1+\left(\frac{u^2\lambda}{(2\alpha)^3}+\frac{3|v|^2\mu}
{2\alpha(\alpha+\beta)^2}
\right)\left(\frac{u^2\lambda}{2\alpha}+\frac{3|v|^2\mu}{\alpha+\beta}
\right)^{-2}\cdot 2|c|^2\right] \ , \qquad\ \ 
\label{5-21}\\
\hbox{\rm (iii)}\quad
F_{cr}&=&
\left(\frac{u^2\lambda}{2\alpha}+\frac{3|v|^2\mu}{2\beta}\right)^{-1} 
\nonumber\\
& &\times 
\left[
1+\left(\frac{u^2\lambda}{(2\alpha)^3}+\frac{3|v|^2\mu}
{(2\beta)^3}
\right)\left(\frac{u^2\lambda}{2\alpha}+\frac{3|v|^2\mu}{2\beta}
\right)^{-2}\cdot 2|c|^2\right] \ , \qquad\ \ 
\label{5-22}
\end{eqnarray}
Under the condition $|\gamma|^2 \sim 0$, the relations (\ref{5-20}) 
$\sim$ (\ref{5-22}) are derived from the relations (\ref{5-4}) $\sim$ 
(\ref{5-6}). 
Further, we have the following forms for the quantity ${\wwtilde f}$ : 
\begin{eqnarray}
\hbox{\rm (i)}\quad
{\wwtilde f}&=&
u^2\sqrt{\frac{\lambda}{\alpha+\beta}}\sqrt{1+\frac{2}{\alpha+\beta}
\left(\lambda-\frac{1}{2\alpha}\right)\frac{|c|^2}
{\frac{u^2\lambda}{\alpha+\beta}+\frac{3|v|^2\mu}{2\alpha}}} \nonumber\\
& &+3|v|^2\sqrt{\frac{\mu}{2\alpha}}\sqrt{1+\frac{1}{\alpha}
\left(\mu-\frac{1}{2\alpha}\right)\frac{|c|^2}
{\frac{u^2\lambda}{\alpha+\beta}+\frac{3|v|^2\mu}{2\alpha}}}
\ , \qquad\ \ 
\label{5-23}\\
\hbox{\rm (ii)}\quad
{\wwtilde f}&=&
u^2\sqrt{\frac{\lambda}{2\alpha}}\sqrt{1+\frac{1}{\alpha}
\left(\lambda-\frac{1}{2\alpha}\right)\frac{|c|^2}
{\frac{u^2\lambda}{2\alpha}+\frac{3|v|^2\mu}{\alpha+\beta}}} \nonumber\\
& &+3|v|^2\sqrt{\frac{\mu}{\alpha+\beta}}\sqrt{1+\frac{2}{\alpha+\beta}
\left(\mu-\frac{1}{2\alpha}\right)\frac{|c|^2}
{\frac{u^2\lambda}{2\alpha}+\frac{3|v|^2\mu}{\alpha+\beta}}}
\ , \qquad\ \ 
\label{5-24}\\
\hbox{\rm (iii)}\quad
{\wwtilde f}&=&
u^2\sqrt{\frac{\lambda}{2\alpha}}\sqrt{1+\frac{1}{\alpha}
\left(\lambda-\frac{1}{2\alpha}\right)\frac{|c|^2}
{\frac{u^2\lambda}{2\alpha}+\frac{3|v|^2\mu}{\alpha+\beta}}} \nonumber\\
& &+3|v|^2\sqrt{\frac{\mu}{2\beta}}\sqrt{1+\frac{1}{\beta}
\left(\mu-\frac{1}{2\beta}\right)\frac{|c|^2}
{\frac{u^2\lambda}{2\alpha}+\frac{3|v|^2\mu}{2\beta}}}
\ , \qquad\ \ 
\label{5-25}
\end{eqnarray}
Under the condition $2|c|^2 \sim 0$, the relations (\ref{5-23}) $\sim$ 
(\ref{5-25}) can be rewritten as 
\begin{eqnarray}
\hbox{\rm (i)}\quad
{\wwtilde f}&=&
\left(u^2\sqrt{\frac{\lambda}{\alpha+\beta}} +3|v|^2\sqrt{\frac{\mu}{2\alpha}}
\right) \nonumber\\
& &\times \sqrt{1+\frac{u^2\left(\sqrt{\frac{\lambda}{\alpha+\beta}}\right)^3
\left(1-\frac{1}{2\alpha\lambda}\right) +3|v|^2\left(\sqrt{\frac{\mu}{2\alpha}}
\right)^3\left(1-\frac{1}{2\alpha\mu}\right)}{
\left(u^2\sqrt{\frac{\lambda}{\alpha+\beta}}+3|v|^2\sqrt{\frac{\mu}{2\alpha}}
\right)\left(\frac{u^2\lambda}{\alpha+\beta}+\frac{3|v|^2\mu}{2\alpha}\right)}
\cdot 2|c|^2} \ , \ \ 
\qquad\ \ 
\label{5-26}\\
\hbox{\rm (ii)}\quad
{\wwtilde f}&=&
\left(u^2\sqrt{\frac{\lambda}{2\alpha}} +3|v|^2\sqrt{\frac{\mu}{\alpha+\beta}}
\right) \nonumber\\
& &\times \sqrt{1+\frac{u^2\left(\sqrt{\frac{\lambda}{2\alpha}}\right)^3
\left(1-\frac{1}{2\alpha\lambda}\right) +3|v|^2\left(\sqrt{\frac{\mu}{\alpha+\beta}}
\right)^3\left(1-\frac{1}{2\alpha\mu}\right)}{
\left(u^2\sqrt{\frac{\lambda}{2\alpha}}+3|v|^2\sqrt{\frac{\mu}{\alpha+\beta}}
\right)\left(\frac{u^2\lambda}{2\alpha}+\frac{3|v|^2\mu}{\alpha+\beta}\right)}
\cdot 2|c|^2} \ , \ \ 
\qquad\ \ 
\label{5-27}\\
\hbox{\rm (iii)}\quad
{\wwtilde f}&=&
\left(u^2\sqrt{\frac{\lambda}{2\alpha}} +3|v|^2\sqrt{\frac{\mu}{2\beta}}
\right) \nonumber\\
& &\times \sqrt{1+\frac{u^2\left(\sqrt{\frac{\lambda}{2\alpha}}\right)^3
\left(1-\frac{1}{2\alpha\lambda}\right) +3|v|^2\left(\sqrt{\frac{\mu}{2\beta}}
\right)^3\left(1-\frac{1}{2\beta\mu}\right)}{
\left(u^2\sqrt{\frac{\lambda}{2\alpha}}+3|v|^2\sqrt{\frac{\mu}{2\beta}}
\right)\left(\frac{u^2\lambda}{2\alpha}+\frac{3|v|^2\mu}{2\beta}\right)}
\cdot 2|c|^2} \ , \ \ 
\qquad\ \ 
\label{5-28}
\end{eqnarray}
Under the form (\ref{5-19}) with the condition $2|c|^2 \sim 0$, 
we can express $(\tau_-)_{cr}$ in the form 
$c\cdot 1/(\sqrt{2}\sqrt{A})\cdot \sqrt{1+B/(2A^2)\cdot 2|c|^2}$. 
In each case of (i), (ii) and (iii), $A$ and $B$ can be given as 
functions of $\alpha$, $\beta$, $\lambda$, $\mu$, $u^2$ and $|v|^2$. 
Compared with the form (\ref{5-19}) for the region 
$2|c|^2 \rightarrow \infty$ is complicatedly related with these parameters.

\section{Discussion}

Starting point of this paper is to give an answer for the problem mentioned 
in \S 6 of (II), in which two approximate forms for the function 
$\tanh |\gamma|^2$ are discussed. 
This function plays a central role in the extension from the conventional 
boson coherent state. 
In the region $2|c|^2 \rightarrow \infty$, the three cases give approximately 
\begin{subequations}\label{6-1}
\begin{eqnarray}
& &(\tau_-)_{cr}=c\sqrt{|c|^2+|v|^2} \ , 
\label{6-1a}\\
& &(\tau_-)_{cr}=c\sqrt{|c|^2+|v|^2} \ , 
\label{6-1b}\\
& &(\tau_-)_{cr}=c\sqrt{|c|^2+(3/2)|v|^2-1/2} \ . 
\label{6-1c}
\end{eqnarray}
\end{subequations}
In the region $2|c|^2 \sim 0$, we have approximate forms : 
\begin{subequations}\label{6-2}
\begin{eqnarray}
(\tau_-)_{cr}&=&c\frac{1}{\sqrt{2}}\frac{1}{\sqrt{u^2+(1/3)|v|^2}} \nonumber\\
& &\times \sqrt{1+(1/3)\frac{u^2+(1/15)|v|^2}{(u^2+(1/3)|v|^2)^2}\cdot
2|c|^2} \ , 
\label{6-2a}\\
(\tau_-)_{cr}&=&c\frac{1}{\sqrt{2}}\frac{1}{\sqrt{u^2+(1/2)|v|^2}} \nonumber\\
& &\times \sqrt{1+(1/2)\frac{u^2+(1/4)|v|^2}{(u^2+(1/2)|v|^2)^2}\cdot
2|c|^2} \ , 
\label{6-2b}\\
(\tau_-)_{cr}&=&c\frac{1}{\sqrt{2}}\frac{1}{\sqrt{u^2+(1/3)|v|^2}} \nonumber\\
& &\times \sqrt{1+(1/3)\frac{u^2+(1/9)|v|^2}{(u^2+(1/3)|v|^2)^2}\cdot
2|c|^2} \ . 
\label{6-2c}
\end{eqnarray}
\end{subequations}
The above $(\tau_-)_{cr}$ is obtained in the form 
\begin{eqnarray}
& &(\tau_-)_{cr}=\gamma^2/2 =c\sqrt{F_{cr}/2} \ , 
\label{6-3}\\
& &|\gamma|^4=2|c|^2\cdot F_{cr}\ . 
\label{6-4}
\end{eqnarray}
Here, $F_{cr}$ is a function of $2|c|^2$. 
The function $F_{cr}$ in each case is obtained through the relation 
\begin{subequations}\label{6-5}
\begin{eqnarray}
& &2|c|^2=|\gamma|^4\left(u^2\frac{\tanh |\gamma|^2}{|\gamma|^2}
+|v|^2\frac{|\gamma|^2\coth |\gamma|^2-1}{|\gamma|^2}\right) \ , 
\label{6-5a}\\
& &2|c|^2=|\gamma|^4\left(u^2\frac{1}{\sqrt{1+|\gamma|^4}}
+|v|^2\frac{1}{\sqrt{1+|\gamma|^4}+1}\right) \ , 
\label{6-5b}\\
& &2|c|^2=|\gamma|^4\left(u^2\frac{1}{\sqrt{9/4+|\gamma|^4}-1/2}
+|v|^2\frac{1}{\sqrt{9/4+|\gamma|^4}+3/2}\right) \ . 
\label{6-5c}
\end{eqnarray}
\end{subequations}
As was already mentioned, there does not exist any state 
which produces the relations (\ref{6-5b}) and (\ref{6-5c}) in the 
framework of the idea of Part (II).

One of the main aim of this section is to investigate the relation 
between the forms presented in \S 5 and those given in this section 
((\ref{6-1}) $\sim$ (\ref{6-5})). 
For this purpose, we must note the function ${\wwtilde f}$ given in the 
relation (\ref{5-18}), which is written as 
\begin{equation}\label{6-6}
(\tau_-)_{cr}=\gamma^2/2\cdot {\wwtilde f} \ . 
\end{equation}
The above is nothing but the relation (\ref{5-19}). 
If omitting the effect of ${\wwtilde f}$ $({\wwtilde f}=1)$, the relation 
(\ref{6-6}) gives, in the region $2|c|^2 \rightarrow \infty$, 
the same forms as those shown in the relation (\ref{6-1b}) and 
(\ref{6-1c}) for the case 
($\alpha=1, \beta=0, \lambda=1, \mu=1/3$) and 
($\alpha=3/2, \beta=-1/2, \lambda=1, \mu=1/3$), respectively. 
But, in any value of the parameters ($\alpha$, $\beta$, $\lambda$, $\mu$), 
the function ${\wwtilde f}$ produces the result (\ref{5-19}). 
Roughly speaking, the result is close to the form (\ref{6-1a}) in a form 
similar to those shown in the relations (\ref{6-1b}) and (\ref{6-1c}).

Our next concern is related to the behavior in the region $2|c|^2 \rightarrow 
0$. If we intend to connect our present form with the relation (\ref{6-2}), 
it may be reasonable to search the form in the following form : 
\begin{eqnarray}\label{6-7}
(\tau_-)_{cr}&=&c\cdot \frac{1}{\sqrt{2}}\cdot 
\frac{1}{\sqrt{A_0u^2+B_0|v|^2}} 
\nonumber\\
& &\times \sqrt{1+\frac{A_1 u^2+B_1|v|^2}{(A_0 u^2+B_0|v|^2)^2}\cdot 2|c|^2} 
\ . 
\end{eqnarray}
Here, we regard $A_0$, $B_0$, $A=1$ and $B_1$ as independent of $u^2$ and 
$|v|^2$. 
With the use of the relations (\ref{5-20}) $\sim$ (\ref{5-22}) and 
(\ref{5-26}) $\sim$ (\ref{5-28}), we can determine the result : 
\begin{equation}\label{6-8}
A_0=1 \ , \quad B_0=1/3 \ , \quad A_1=1 \ , \quad B_1=1/27 \ . 
\end{equation}
It should be noted that the result (\ref{6-8}) is common to the three cases. 
Then, we have 
\begin{eqnarray}\label{6-9}
(\tau_-)_{cr}&=&c\cdot \frac{1}{\sqrt{2}}\cdot 
\frac{1}{\sqrt{u^2+(1/3)|v|^2}} 
\nonumber\\
& &\times \sqrt{1+\frac{u^2+(1/27)|v|^2}{(u^2+(1/3)|v|^2)^2}\cdot 2|c|^2} 
\ . 
\end{eqnarray}
However, the condition producing the result (\ref{6-8}) is not common 
to the three cases : 
\begin{subequations}\label{6-10}
\begin{eqnarray}
\hbox{\rm (i)}\quad
{\lambda}&=&(9/16)\frac{\alpha-\beta}{\alpha\cdot(2\alpha)} \ , \qquad
\mu=(1/16)\frac{\alpha-\beta}{\alpha\cdot(\alpha+\beta)} \ , \quad 
(\alpha > \beta) \qquad\qquad\qquad
\label{6-10a}\\
\hbox{\rm (ii)}\quad
{\lambda}&=&(9/16)\frac{\beta-\alpha}{\alpha\cdot(2\alpha)} \ , \qquad
\mu=(1/16)\frac{\beta-\alpha}{\alpha\cdot(\alpha+\beta)} \ , \quad 
(\alpha < \beta) \qquad\qquad\qquad
\label{6-10b}\\
\hbox{\rm (iii)}\quad
{\lambda}&=&(9/16)\left(\frac{1}{\alpha}+\frac{1}{\beta}\right)^2/2\cdot
(\beta-\alpha) \ , \nonumber\\
{\mu}&=&(1/16)\left(\frac{1}{\alpha}+\frac{1}{\beta}\right)^2/2\cdot
2\alpha(\beta-\alpha) \ , \quad 
(\alpha < \beta) \qquad\qquad\qquad
\label{6-10c}
\end{eqnarray}
\end{subequations}
Comparison of the result (\ref{6-9}) with the form (\ref{6-2}) 
may be interesting. The leading term is completely the same as each other. 
But, the term in the next order is larger than that in the form (\ref{6-2}). 
This comes from the effect of ${wwtilde f}$.

Our next interest is to investigate the expectation value of ${\hat c}$, which 
we denote as $(c)_{cr}$. In this case, the formula shown in the relation 
(II.2$\cdot$33) is also available : 
\begin{eqnarray}\label{6-11}
(c)_{cr}&=&uv|\gamma|\sqrt{\Gamma_{ch}/\Gamma_{sh}} 
\left({\wtilde f}(N)\right)_{ch} \nonumber\\
& &+uv^*(\gamma^2/|\gamma|)\sqrt{\Gamma_{ch}/\Gamma_{sh}} 
\left({\wtilde f}(N)\right)_{sh}  \ . 
\end{eqnarray}
In the state (\ref{3-1}), which we are mainly interested in, we have 
\begin{subequations}\label{6-12}
\begin{eqnarray}
& &\left({\wtilde f}(N)\right)_{ch}=
\frac{\sqrt{1-|\gamma|^4 g_{ch}^2}}{\left(\sqrt{1-|\gamma|^4 g_{ch}g_{sh}}
\right)^3} \ , 
\label{6-12a}\\
& &\left({\wtilde f}(N)\right)_{sh}=g_{ch}
\frac{\left(\sqrt{1-|\gamma|^4 g_{sh}^2}\right)^3}
{\left(\sqrt{1-|\gamma|^4 g_{ch}g_{sh}}\right)^3} \ . 
\label{6-12b}
\end{eqnarray}
\end{subequations}
The quantities $\Gamma_{ch}$ and $\Gamma_{sh}$ are given in the relation 
(\ref{3-4}). 
Then, we have the following form for $(c)_{cr}$ : 
\begin{eqnarray}\label{6-13}
(c)_{cr}&=&(uv+uv^*\gamma^2 g_{ch})
\frac{\sqrt{\sqrt{1-|\gamma|^4 g_{ch}^2}\left(\sqrt{1-|\gamma|^4g_{ch}^2}
\right)^3}}{\left(\sqrt{1-|\gamma|^4g_{ch}g_{sh}}\right)^3} \ . 
\end{eqnarray}
In the case (\ref{4-1}), together with the relation (\ref{4-3}), 
$(c)_{cr}$ is expressed in the form 
\begin{eqnarray}\label{6-14}
(c)_{cr}&=&\frac{u}{\sqrt{\frac{u^2+3|v|^2\mu/\lambda}{2}}}
\left(v\sqrt{\frac{u^2+3|v|^2\mu/\lambda}{2}+|c|^2}+v^*c\right)
\nonumber\\
& &\times \left[
\sqrt{\frac{\sqrt{(u^2+3|v|^2\mu/\lambda+2|c|^2)
(u^2+3|v|^2\mu/\lambda+2|c|^2\mu/\lambda)}+\sqrt{\mu/\lambda}\cdot 2|c|^2}
{u^2+3|v|^2\mu/\lambda+(1+\mu/\lambda)\cdot 2|c|^2}}\right]^3 \ , 
\nonumber\\
u^2&=&1-|v|^2 \ , \qquad v=\eta \ . 
\end{eqnarray}
In the case $\lambda=\mu$, we have 
\begin{equation}\label{6-15}
(c)_{cr}=\sqrt{\frac{1-|v|^2}{1/2+|v|^2}}\left(
\sqrt{1/2+|v|^2+|c|^2}+cv^*\right) \ . 
\end{equation}
The above is of the same form as that obtained in Ref.\citen{four}. 
Further, in the case $\lambda=3\mu$, the form (\ref{6-14}) is 
reduced to 
\begin{eqnarray}\label{6-16}
(c)_{cr}&=&\sqrt{2}{u}\times \left(\sqrt{\frac{\sqrt{3}}{4}}\right)^3
\left(\sqrt{1/2+|c|^2}v+\frac{c}{\sqrt{3}}v^*\right) 
\nonumber\\
& &\times \left(
\sqrt{\frac{\sqrt{(1/2+|c|^2)
(3/2+|c|^2)}+|c|^2}{3/8+|c|^2}}\right)^3 \ . 
\end{eqnarray}
The cases (\ref{5-1}) $\sim$ (\ref{5-3}) with $\alpha=\beta$ are reduced to 
the same forms as that shown in the relation (\ref{6-16}). 
This is the same situation as that in the relations (\ref{4-4}) and 
(\ref{5-15}) for $(\tau_-)_{cr}$. 
The other cases are also in the situation similar to the cases 
$(\tau_-)_{cr}$ already discussed.

\acknowledgement

Main part of the investigation presented in this paper was performed 
when one of the authors (M. Y.) stayed at Coimbra. in summer of 2003. 
He expresses his sincere thanks to Professor J. da Provid\^encia, 
co-author of this paper, for his kind invitation to Coimbra.

\end{document}